\documentclass[secnumarabic, graphics,floatfix, nofootinbib,tightenlines,nobibnotes, aps, prl, 12pt]{revtex4-1}
\usepackage{graphicx}
\usepackage[english]{babel}
\usepackage[utf8]{inputenc}
\usepackage{amsmath,amssymb}
\usepackage{amsfonts}
\usepackage{multirow}
\usepackage{pstricks}
\usepackage{float}
\usepackage{subfigure}
\usepackage{color}
\usepackage{epsfig}
\usepackage{pst-tools}
\usepackage{mnsymbol}
\usepackage[section]{placeins}

\begin{document}

\title{On the peripheral-tube description of the two-particle correlations in nuclear collisions}

\author{Dan Wen$^1$}
\author{Wagner Maciel Castilho$^1$}
\author{Kai Lin$^{2,3}$}
\author{Wei-Liang Qian$^{3,1,4}$}
\author{Yogiro Hama$^{5}$}
\author{Takeshi Kodama$^{6,7}$}

\affiliation{$^{1}$ Faculdade de Engenharia de Guaratinguet\'a, Universidade Estadual Paulista, 12516-410, Guaratinguet\'a, SP, Brazil}
\affiliation{$^{2}$ Escola de Engenharia de Lorena, Universidade de S\~ao Paulo, 12602-810, Lorena, SP, Brazil}
\affiliation{$^{3}$ Hubei Subsurface Multi-scale Imaging Key Laboratory, Institute of Geophysics and Geomatics, China University of Geosciences, 430074, Wuhan, Hubei, China}
\affiliation{$^{4}$ School of Physical Science and Technology, Yangzhou University, 225002, Yangzhou, Jiangsu, China}
\affiliation{$^{5}$ Instituto de F\'isica, Universidade de S\~ao Paulo, C.P. 66318, 05315-970, S\~ao Paulo, SP, Brazil}
\affiliation{$^{6}$ Instituto de F\'isica, Universidade Federal do Rio de Janeiro, C.P. 68528, 21945-970, Rio de Janeiro, RJ , Brazil}
\affiliation{$^{7}$ Instituto de F\'isica, Universidade Federal Fluminense, 24210-346, Niter\'oi, RJ, Brazil}

\date{Sept. 30, 2018}

\begin{abstract}

In this work, we study the two-particle correlations regarding a peripheral tube model.
From our perspective, the main characteristics of the observed two-particle correlations are attributed to the multiplicity fluctuations and the locally disturbed one-particle distribution associated with hydrodynamic response to the geometric fluctuations in the initial conditions.
We investigate the properties of the initial conditions and collective flow concerning the proposed model.
It is shown that the experimental data can be reproduced by hydrodynamical simulations using appropriately constructed initial conditions. 
Besides, instead of numerical calibration, we extract the model parameters according to their respective physical interpretations and show that the obtained numerical values are indeed qualitatively in agreement with the observed data.
Possible implications of the present approach are discussed.


\end{abstract}

\maketitle

\section{I. Introduction}

Event-by-event fluctuations play a crucial role in the hydrodynamic description of the relativistic heavy ion collisions~\cite{sph-review-1,hydro-review-04,hydro-review-05,hydro-review-06,hydro-review-09,hydro-review-10}. 
The experimental observations of the enhancement in correlations at intermediate and low transverse momenta~\cite{RHIC-star-ridge-3,RHIC-phenix-ridge-5,RHIC-phobos-ridge-7,LHC-alice-vn-4,LHC-cms-ridge-7,LHC-atlas-vn-4}, comparing with those at high transverse momenta~\cite{RHIC-star-jet-1,RHIC-star-jet-2} strongly support the hydrodynamics/transport characteristic of such phenomena~\cite{sph-corr-1,hydro-v3-1,hydro-vn-ph-3,hydro-corr-ph-1,hydro-corr-1,ampt-4,ampt-5}.
In particular, the observed features, referred to as ``ridge'' and ``shoulders'', were shown to be successfully reproduced by hydrodynamical simulations with event-by-event fluctuating initial conditions (IC) but not by averaged IC~\cite{sph-corr-1}.
Subsequently, it leads to the current understanding through extensive studies of event-by-event based hydrodynamic/transport analysis~\cite{hydro-v3-2,hydro-v3-4,hydro-v3-5,hydro-v3-6,hydro-v3-7,hydro-v3-8}, that the two-particle correlations for the lower transverse momenta can be mostly interpreted in terms of flow harmonics $v_n$.
Notably, the triangular flow, $v_3$, is mostly attributed to for the appearance of the ``shoulder" structure on the away side of the trigger particle~\cite{hydro-v3-1,hydro-v3-2,hydro-v3-3}.
Moreover, it is understood that these parameters are closely associated with the corresponding $\epsilon_n$, the anisotropies of the initial energy distribution~\cite{hydro-v3-1,hydro-v3-2}.
In fact, the observed behavior of anisotropic parameters as functions of centrality and transverse momentum has been studied at a very high quantitative level~\cite{hydro-v2-heinz-2,hydro-v2-voloshin-1}, as well as the correlation between $v_{n}$ and $\epsilon_n$ has also been established (see~\cite{hydro-review-08,hydro-review-04,hydro-review-09,hydro-review-06} for the recent reviews on this topic).

In spite of the success of the statistical analysis of event-by-event hydrodynamic simulations, the linearity between event averaged  $v_n$ and $\epsilon_n$ become less evident for harmonics greater than $n=2$. 
To be specific, it was shown that the correlations among $v_n$ and $\epsilon_n$ become weaker for larger harmonics~\cite{hydro-vn-3,sph-vn-4}.  
This suggests that the event-by-event fluctuations themselves carry important information.  
Subsequently, if one restricts himself only to the analysis of the event-averaged relations/correlations among  $v_n$ and $ \epsilon_n$, then some genuine hydrodynamic signals from the local fluctuations in each individual event might be washed out, or hidden behind some very complicated correlations among the harmonics~\cite{hydro-corr-ph-7,hydro-corr-ph-8,hydro-vn-10}.  
That is, the local genuine hydrodynamic evolution, e.g., turbulence, local shock wave, etc., while sensitive to the properties of matter, might not be encoded in a simple way regarding event-average correlations.  
In this context, in the present paper, we explore from an alternative angle which may explain in a simple way the physical origin of the anisotropic flow pattern.

Before the explanation regarding the triangular flow was first suggested~\cite{hydro-v3-1}, we proposed a peripheral tube model~\cite{sph-corr-2,sph-corr-3,sph-corr-4,sph-corr-7} which provides a straightforward and reasonable picture for the generation of the triangular flow and the consequent two-particle correlations.
It is an approach within the general event-by-event hydrodynamic scheme. 
The model views the fluctuations in the IC as consisting of independent high energy tubes close to the surface of the system.
Thereby, each tube affects the hydrodynamical evolution of the system independently, and their contributions are summed up linearly to the resultant two-particle correlations. 
In this approach, one substitutes the complex bulk of the hot matter by an average distribution over many events from the same centrality class.

The above picture attempts to interpret the physical content of fluctuating IC regarding the granularity represented by peripheral high energy tubes.
To be specific, if a tube locates deep inside the hot matter, the effect of its hydrodynamic expansion would be quickly absorbed by its surroundings, causing relatively less inhomogeneity in the media.
On the contrary, a tube staying close to the surface leads to a significant disturbance to the one-particle distribution, resulting in an azimuthal two-particle correlation structure similar in shape and magnitude to the observed data.

\begin{figure}
\begin{tabular}{cc}
\vspace{-50pt}
\begin{minipage}{225pt}
\centerline{\includegraphics[width=350pt]{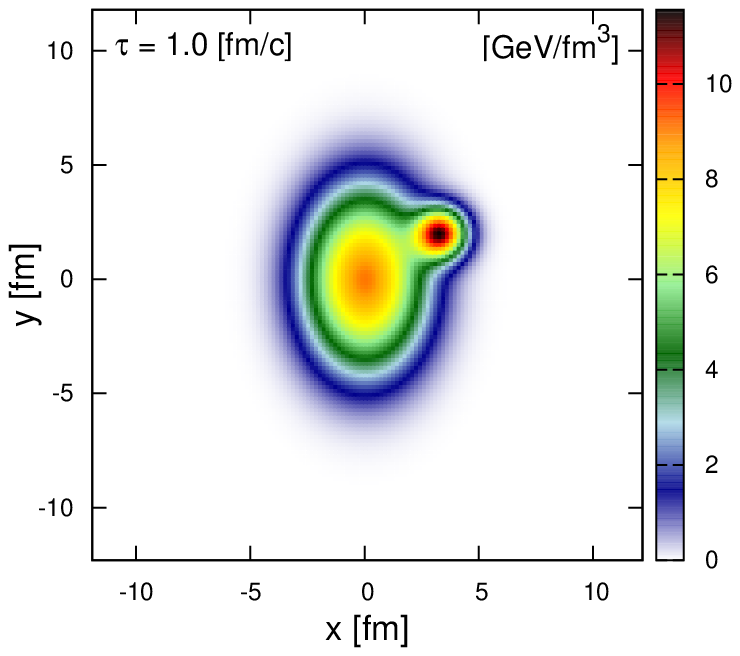}}
\end{minipage}
&
\begin{minipage}{225pt}
\centerline{\includegraphics[width=350pt]{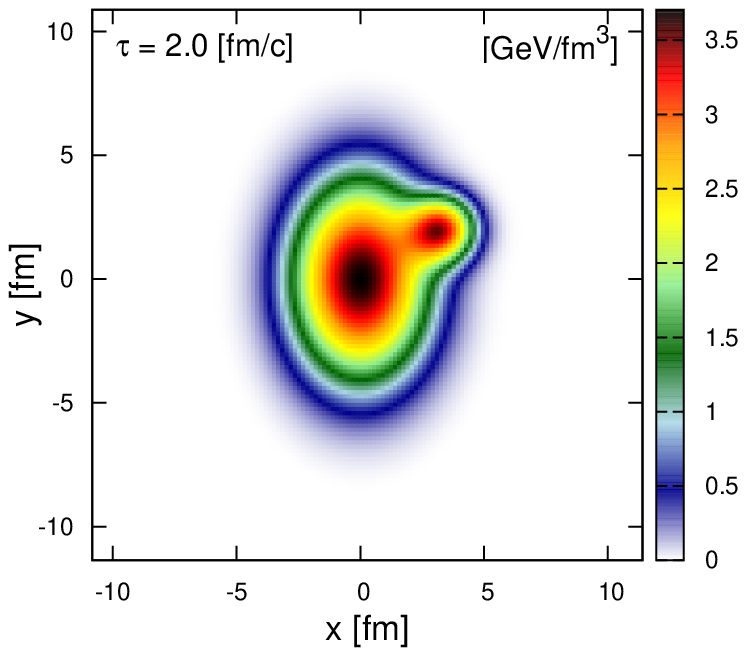}}
\end{minipage}
\\
\vspace{-50pt}
\begin{minipage}{225pt}
\centerline{\includegraphics[width=350pt]{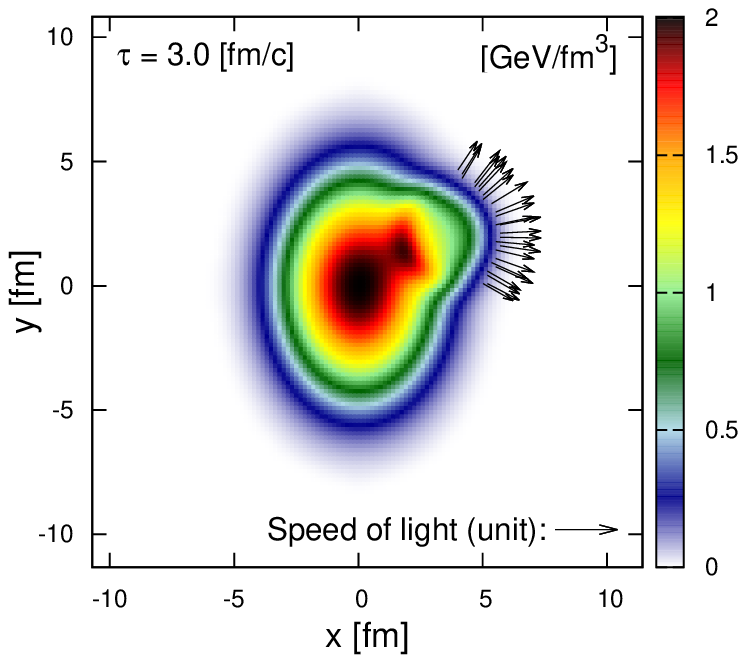}}
\end{minipage}
&
\begin{minipage}{225pt}
\centerline{\includegraphics[width=350pt]{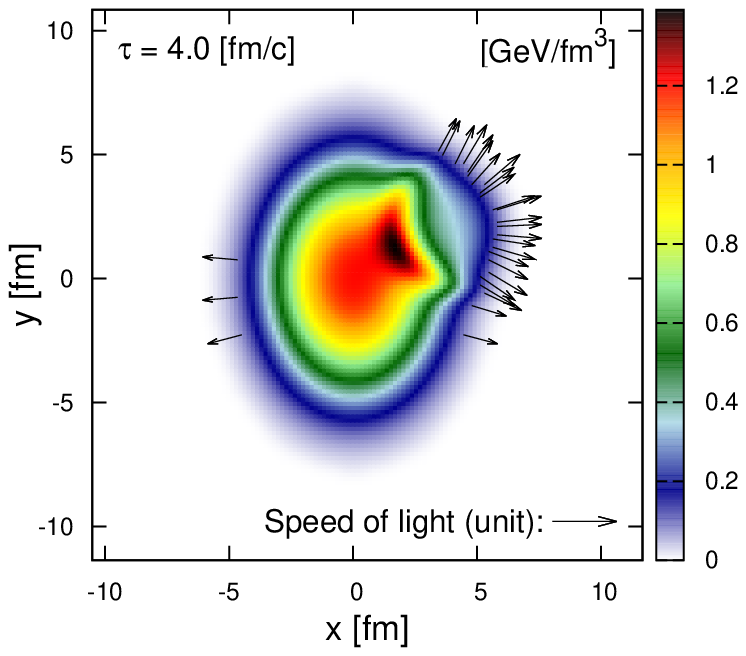}}
\end{minipage}
\\
\begin{minipage}{225pt}
\centerline{\includegraphics[width=350pt]{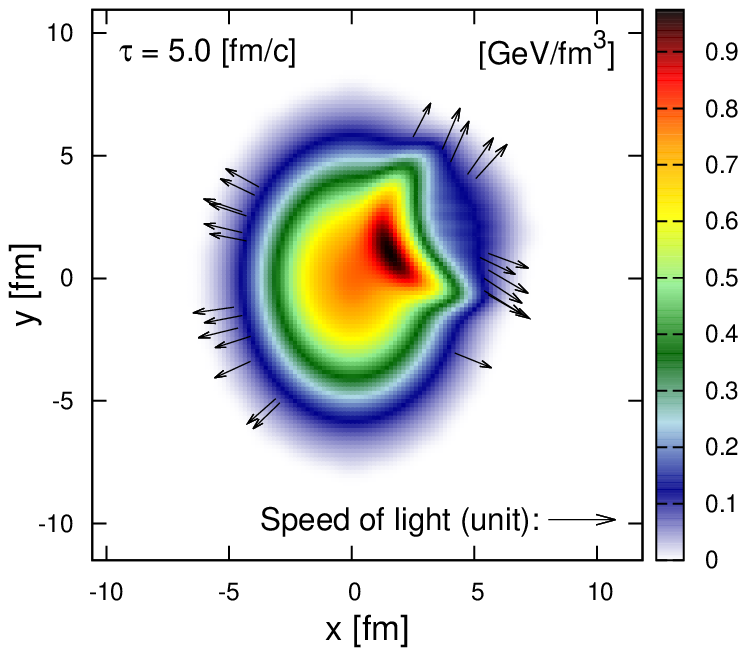}}
\end{minipage}
&
\begin{minipage}{225pt}
\centerline{\includegraphics[width=350pt]{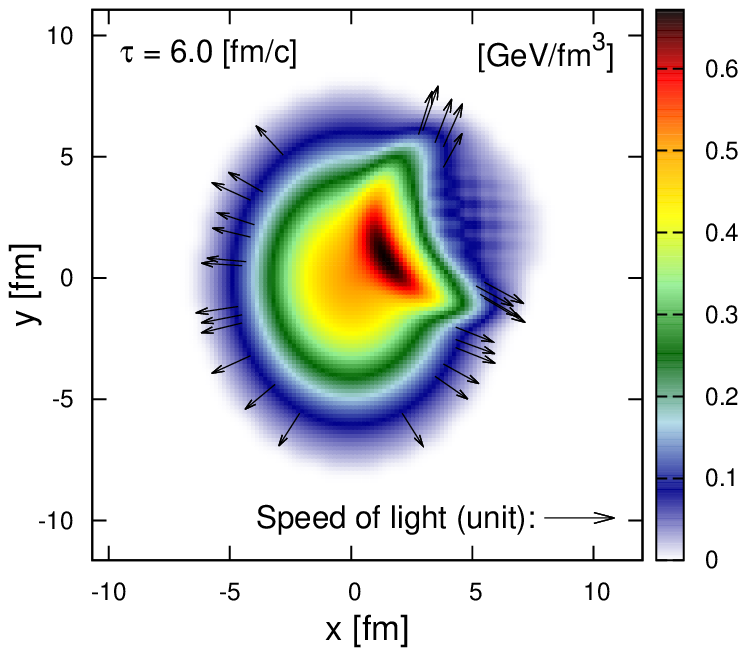}}
\end{minipage}
\end{tabular}
\caption{(Color online) The temporal evolution of the IC consisting of one peripheral tube placed on top of an elliptic smoothed background energy distribution. 
The parameters for the IC used in calculation are discussed in section III.}
 \label{figevo}
\end{figure}

By numerical simulations, as shown in Fig.\ref{figevo}, one finds that the fluid is deflected to both sides of the tube, causing two peaks separated by $\sim 120$ degrees in the one-particle azimuthal distribution.
Subsequently, this leads to the desired two-particle distribution where a double peak is formed on the away side, whose height is approximately half of the single peak located on the near side. 
It is shown that the resultant correlation structure is robust against the variation of the model parameters~\cite{sph-corr-2,sph-corr-4}.
Furthermore, simulations carried out with multiple peripheral tubes still show such robust feature of the two-particle correlation structure, which strongly suggests that the emergence of the two particle correlations can be naturally interpreted as a superposition of those of independent peripheral tubes~\cite{sph-corr-7}.

It is interesting to compare the present model to the interpretation frequently employed in the literature, where random but systematic emergence of octupole deformation in the initial energy distribution is considered to lead to the final triangular flow, and consequently, the double peak nature of the two particle correlations in the final state.  
The mostly linear relation between the initial anisotropy parameters, $\epsilon_n$ and the flow harmonics $v_n$, observed in numerical simulations~\cite{hydro-vn-2,hydro-vn-3,hydro-vn-4,hydro-vn-9} are considered as the evidence of such interpretation.  
We note that it is meaningful to compare the present model to another approach frequently employed in the literature, where the one-particle distribution are decomposed into different flow harmonics.
Subsequently, the collective flow related to the harmonic coefficients are understood to be mostly independent and associated with the corresponding eccentricity components in the IC.
Note that the physical image associated with the above interpretation is drastically different from that of our approach regarding the emergence of octupole moment of the initial energy distribution. 
In our model, the octupole moment does appear in IC, but it is not directly related to the formation of the triangular flow. 
In both cases, by and large, the hydrodynamic evolution linearly transforms the initial state geometric inhomogeneity into the final state anisotropy in momentum space.
The main point is that in our present model the effect is considered as local, as a result of how the expansion is affected by a high energy tube close to the surface of the fluid.
Therefore, it is irrelevant to the fluid dynamics of the rest of the system, but localized to a specific azimuthal angle $\phi_\mathrm{tube}$ associated to the peripheral tube in question.
As hydrodynamics is understood as an effective theory at the long wavelength limit, the peripheral tube interprets the two-particle correlation in terms of phenomena where the characteristic length is comparable to the system size.
Concerning harmonic coefficients, the event planes of the elliptic and triangular flow coefficients are both generated by the tube and are therefore both correlated to the location of the tube $\phi_\mathrm{tube}$.
However, as one carries out the event-by-event average, $\phi_\mathrm{tube}$ is averaged out, and the resultant expression (for instance, see Eqs.(\ref{ccumulant}) and (\ref{ccumulant2}) below) does not explicitly depend on it.

The present work is organized as follows.
In the next section, we briefly review the peripheral tube model and discuss its main results on the observed two-particle correlations.
Subsequently, in section III, we show that the experimental data can be reasonably reproduced by appropriately constructing the IC with the peripheral tube.
Furthermore, we extract the model parameters using event-by-event simulations with various devised IC reflecting the average background and event-by-event fluctuations.
It is shown that those extracted values are indeed in accordance with the observed two-particle correlation data.
The last section is devoted to discussions and concluding remarks.

\section{II. The peripheral tube model for two-particle correlations}

The underlying assumption of the peripheral tube model is that the hydrodynamic evolution which generates the characteristic double-peaked two-particle correlation is mainly composed of two parts.
The first one is associated with the bulk energy distribution of the IC, which generates the dominant part of the resultant collective flow.
For simplicity, this bulk energy distribution is assumed to generate mostly the elliptic flow.  
Owing to the multiplicity fluctuations, the contributions from the proper events does not cancel out with those from the mixed event, and the residual is proportional to the standard deviation of the multiplicities.
The second contribution comes from that of a peripheral tube.
It produces aligned elliptic and triangular flow.
This fact is related to the event plane dependence as well as centrality dependence of the observed two-particle correlation~\cite{sph-corr-ev-4,sph-corr-ev-6}.
To be specific, in this model, the two-particle correlation is entirely determined by the one-particle distribution.
Instead of writing the latter down directly in terms of Fourier expansion~\cite{hydro-corr-ph-2}, we write down the one-particle distribution as a sum of two terms, namely, the distribution of the background and that of the tube.
\begin{eqnarray} 
  \frac{dN}{d\phi}(\phi,\phi_\mathrm{tube}) =\frac{dN_{\mathrm{bgd}}}{d\phi}(\phi) +\frac{dN_{\mathrm{tube}}}{d\phi}(\phi,\phi_\mathrm{tube}), 
 \label{eq-sec5-1-1}  
\end{eqnarray} 
where 
\begin{eqnarray} 
 \frac{dN_{\mathrm{bgd}}}{d\phi}(\phi)&=&\frac{N_\mathrm{bgd}}{2\pi}(1+2v_2^\mathrm{bgd}\cos(2\phi)),    \label{eq-sec5-1-2}\\  
 \frac{dN_{\mathrm{tube}}}{d\phi}(\phi,\phi_\mathrm{tube})&=&\frac{N_\mathrm{tube}}{2\pi}\sum_{n=2,3}2v_n^\mathrm{tube}\cos(n[\phi-\phi_\mathrm{tube}])  . \label{eq-sec5-1-3}
\end{eqnarray} 
In Eq.(\ref{eq-sec5-1-2}) we consider the most simple case for the background flow, by parametrizing it with the elliptic flow parameter $v_2^\mathrm{bgd}$ and the overall multiplicity, denoted by $N_\mathrm{bgd}$. 
The contributions from the tube are measured with respect to its angular position $\phi_\mathrm{tube}$, and a minimal number of Fourier components are introduced to reproduce the desired two-particle correlation~\cite{sph-corr-4}, that is to say, only $v_2^\mathrm{tube}$ and $v_3^\mathrm{tube}$ are retained in Eq.(\ref{eq-sec5-1-3}). 
It is worth noting that although both the contributions from the background and the tube are written in Fourier expansion, they are intrinsically independent distributions. 
In particular, the triangular flow in our model is completely generated by the tube, and since its symmetry axis is correlated to the tube location $\phi_\mathrm{tube}$, the variation of the latter is related to the event-by-event fluctuations.
We also assume that the flow components from the background are much more significant than those generated by the tube, $\Psi_2$ is mainly determined by the elliptic flow of the background $v_2^\mathrm{bgd}$. 

Following the methods used by the STAR experiment~\cite{RHIC-star-plane-1,RHIC-star-plane-2,RHIC-star-plane-3}, the subtracted di-hadron correlation is given by
\begin{eqnarray} 
 \left<\frac{dN_{\mathrm{pair}}}{d\Delta\phi}(\phi_s)\right>   =\left<\frac{dN_{\mathrm{pair}}}{d\Delta\phi}(\phi_s)\right>^{\mathrm{prop}} -\left<\frac{dN_{\mathrm{pair}}}{d\Delta\phi}(\phi_s)\right>^{\mathrm{mix}} , \nonumber \\
 \label{eq-sec5-1-4}  
\end{eqnarray} 
where $\phi_s$ is the trigger angle ($\phi_s=0$ for in-plane and $\phi_s=\pi/2$ for out-of-plane trigger). 
By using Eq.(\ref{eq-sec5-1-3}), one finds the proper two-particle correlation
\begin{eqnarray} 
\left<\frac{dN_{\mathrm{pair}}}{d\Delta\phi}\right>^{\mathrm{prop}}  =
 \int\frac{d\phi_\mathrm{tube}}{2\pi}f(\phi_\mathrm{tube})
 \frac{dN^T}{d\phi}(\phi_s,\phi_\mathrm{tube})  
\frac{dN^A}{d\phi} (\phi_s+\Delta\phi,\phi_\mathrm{tube}),   \nonumber \\ \label{eq-sec5-1-corr-proper}
\end{eqnarray}
where $f(\phi_\mathrm{tube})$ is the distribution function of the tube, and superscripts ``$T$" and ``$A$" indicate ``trigger" and ``associated" particles respectively (c.f. subscripts ``$T$" are shorthands for ``transverse").
For simplicity, we take $f(\phi_\mathrm{tube})=1$. 
In a more realistic approach, however, $f(\phi_\mathrm{tube})$ might contain a small elliptic modulation owing to the elliptic shape of the IC.
Nonetheless, as can be shown straightforwardly, such correction leads to additional contributions which are smaller than the results discussed below, and therefore are ignored in the present study.

The combinatorial background $\left<{dN_{\mathrm{pair}}}/{d\Delta\phi}\right>^{\mathrm{mix}}$ can be calculated by using either cumulant or ZYAM method \cite{zyam-1}. 
Though both methods yield very similar results in our model, it is more illustrative to evaluate the cumulant. 
Following similar arguments presented in Ref.~\cite{sph-corr-ev-4}, it is straightforward to show that the resultant correlation reads
\begin{eqnarray}
\langle \frac{dN_\mathrm{pair}}{d\Delta\phi}(\phi_s)\rangle ^{(\mathrm{cmlt})} 
 &=&\frac{\langle N_\mathrm{bgd}^T N_\mathrm{bgd}^A\rangle -\langle N_\mathrm{bgd}^T\rangle \langle N_\mathrm{bgd}^A\rangle }{(2\pi)^2}(1+2v_2^{\mathrm{bgd},T}\cos(2\phi_s))
 (1+2v_2^{\mathrm{bgd},A}\cos(2(\Delta\phi+\phi_s))) \nonumber\\  
 &+&\frac{\langle N_\mathrm{tube}^T N_\mathrm{tube}^A\rangle }{(2\pi)^2}\sum_{n=2,3}2v_n^{\mathrm{tube},T}v_n^{\mathrm{tube},A}
 \cos(n\Delta\phi) \ , 
 \label{ccumulant}
\end{eqnarray} 
where the event average is carried out by integration in $\phi_\mathrm{tube}$. 
The above expression explicitly depends on $\phi_s$, which can be used to study the event plane dependence of the correlation.
In particular, the trigonometric dependence of the background contribution on $\phi_s$ indicates that its contribution to the out-of-plane triggers is opposite to that for the in-plane ones. 
As a result, for the out-of-plane correlation, it leads to an overall suppression in the amplitude, as well as forms a double peak structure on the away side.
Indeed, experimental data~\cite{RHIC-star-plane-1,RHIC-star-plane-2,RHIC-star-plane-3} shows that the overall correlation decreases while the away-side correlation evolves from a single peak to a double peak as $\phi_s$ increases. 
Since these observed features are in agreement with the analytically derived results, the peripheral model is shown to be meaningful despite of its simplicity.

By further averaging out $\phi_s$, one obtains
\begin{eqnarray}
\langle \frac{dN_\mathrm{pair}}{d\Delta\phi}\rangle ^{(\mathrm{cmlt})} 
 &=&\frac{\langle N_\mathrm{bgd}^T N_\mathrm{bgd}^A\rangle -\langle N_\mathrm{bgd}^T\rangle \langle N_\mathrm{bgd}^A\rangle }{(2\pi)^2}(1+2v_2^{\mathrm{bgd},T}v_2^{\mathrm{bgd},A}\cos(2\Delta\phi)) \nonumber\\  
 &+&\frac{\langle N_\mathrm{tube}^T N_\mathrm{tube}^A\rangle }{(2\pi)^2}\sum_{n=2,3}2v_n^{\mathrm{tube},T}v_n^{\mathrm{tube},A}
 \cos(n\Delta\phi). 
 \label{ccumulant2}
\end{eqnarray}
If the model is indeed realistic, one should be able to obtain the parameters in Eq.(\ref{ccumulant2}) according to their respective physical content, while the resulting correlations should be still quanlitatively in agreement with the data.
This is the principal object of the present work.
In what follows, by carrying out numerical simulations, we first show that an appropriately constructed IC can reasonably reproduce the observed data. 
Furthermore, we attempt to calculate the model parameters according to their definitions.
This is done by using various IC tailored to match the respective physical properties of IC in question.
In particular, we study the multiplicity fluctuations as well as the flow harmonics of different IC associated with the background as well as the peripheral tube.
The two-particle correlations are then evaluated by using the obtained values.

\begin{figure}
\begin{tabular}{cc}
\begin{minipage}{225pt}
\centerline{\includegraphics[width=200pt]{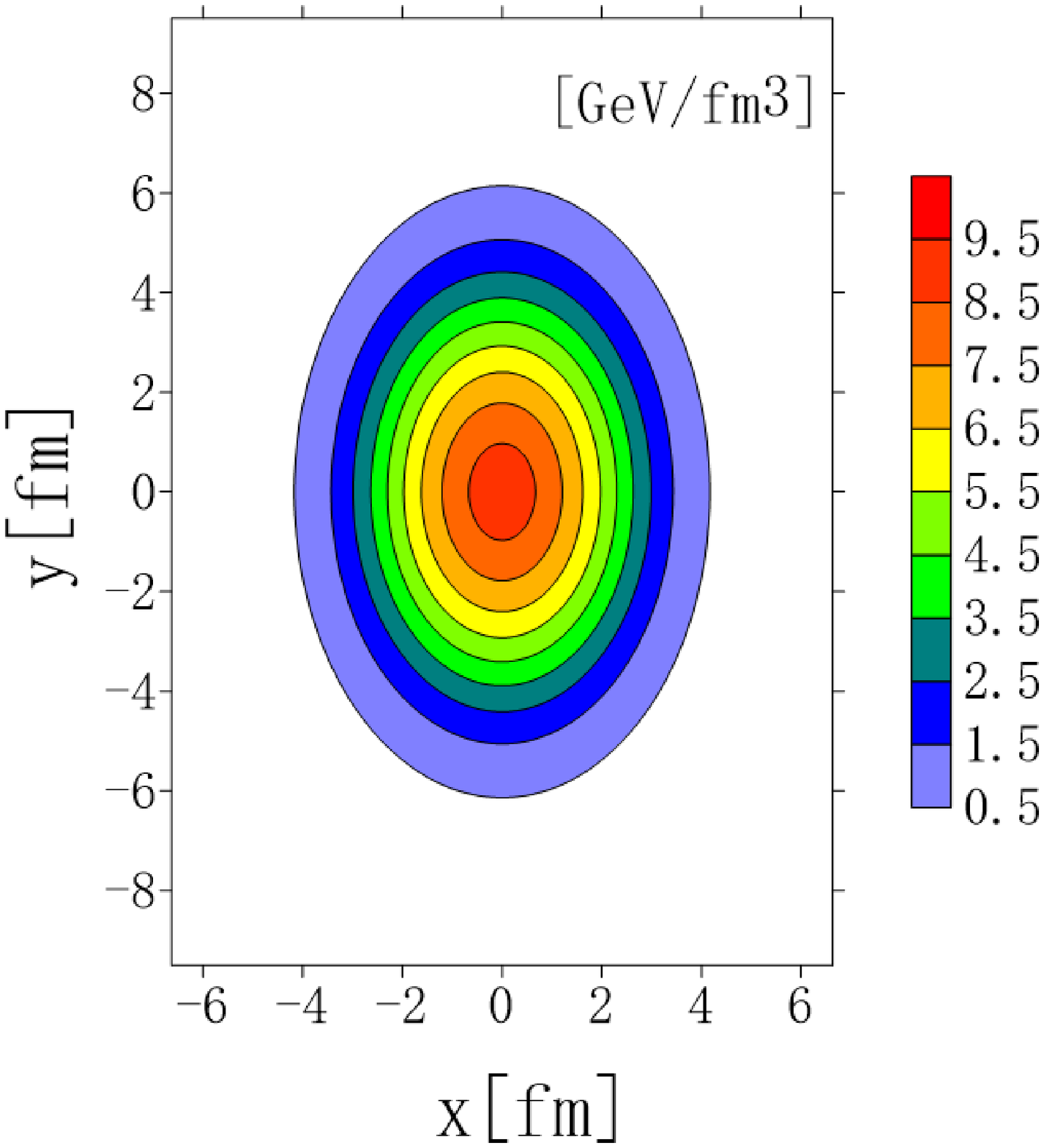}}
\end{minipage}
&
\begin{minipage}{225pt}
\centerline{\includegraphics[width=200pt]{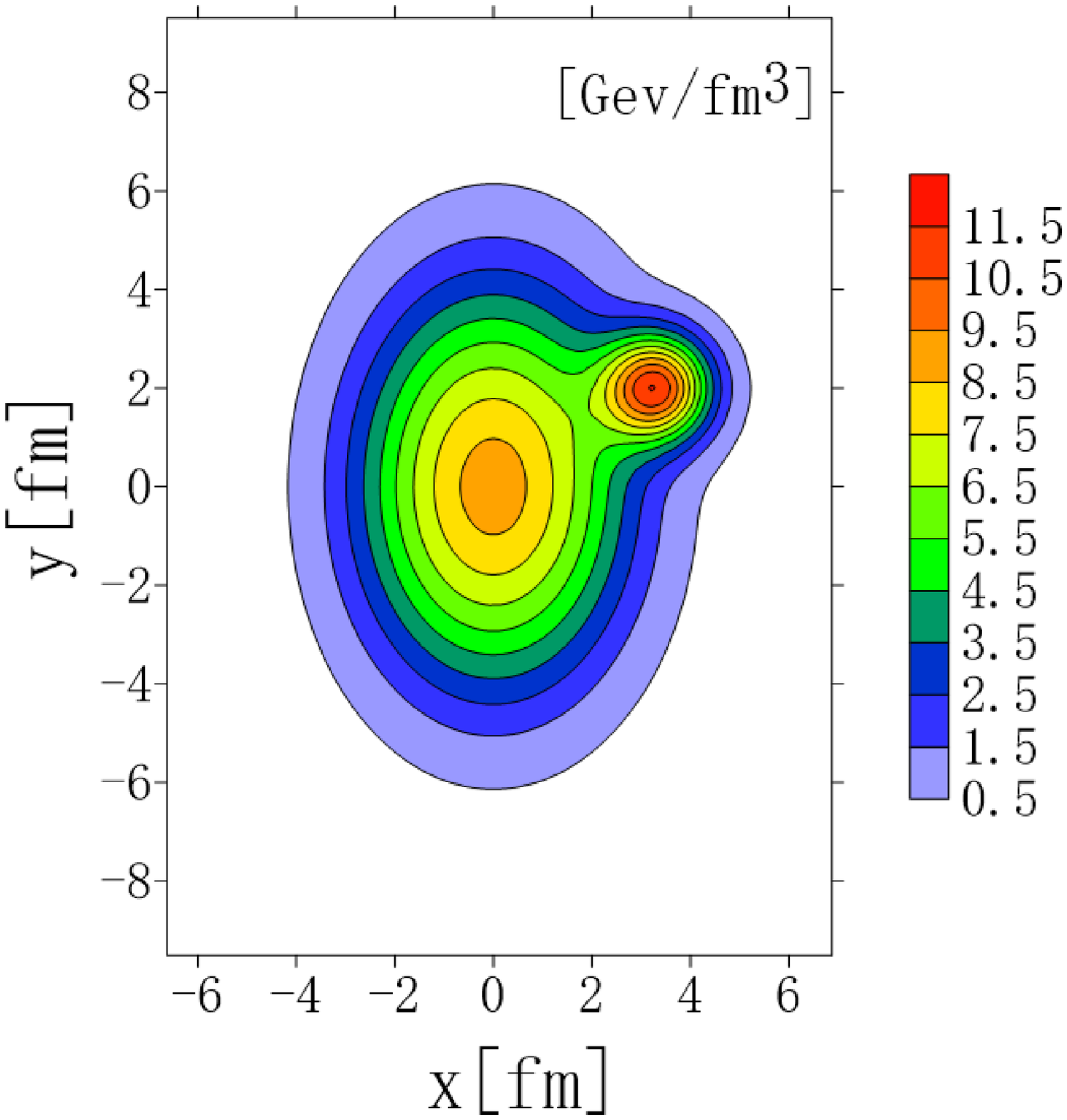}}
\end{minipage}
\end{tabular}
\caption{(Color online) Left: the background energy distribution obtained by averaging over many NeXuS events. Right: one random event with a high energy peripheral tube sits on top of the background.
The parametrization of the IC is discussed in the text.}
 \label{figIC}
\end{figure}

\subsection{III. The model parameters extracted by comparing hydrodynamic simulations against the data}

For the present study, we focus on mid-central 200 AGeV Au+Au collisions.
We first show that one can devise the IC according to the peripheral tube model and subsequently evaluate the two-particle correlation.
This is done by estimating the background energy distribution by averaging over the 343 events generated by a microscopic event generator, NeXuS~\cite{nexus-rept}, for the centrality window 20\%-40\% as shown in Fig.\ref{figIC}.
The obtained almond shaped energy distribution is then fitted by the following parametrization
\begin{eqnarray}
\epsilon_\mathrm{bgd} &=& (9.33+7r^2+2r^4)e^{-r^{1.8}}  \ \ \text{fm}^{-3}\, , \label{energyb} 
\end{eqnarray}
with
\begin{eqnarray}
r &=& \sqrt{0.41x^2+0.186y^2} \ \ \text{fm} .   \nonumber
\end{eqnarray}

The profile of the high energy tube is calibrated to that of a typical peripheral tube in NeXuS IC as follows
\begin{eqnarray}
\epsilon_\mathrm{tube}&=& 12e^{-(x-x_\mathrm{tube})^2-(y-y_\mathrm{tube})^2} \ \ \text{fm}^{-3} \, , \label{energyt}
\end{eqnarray}
where the tube is located at a given value of energy density close to the surface, determined by a free parameter $r_\mathrm{tube}$, so that its coordinates on the transverse plane read
\begin{eqnarray}
x_\mathrm{tube} &=& \frac{r_\mathrm{tube}\cos\theta}{\sqrt{0.41\cos^2\theta+0.186\sin^2\theta}}   \ \ \text{fm}\\
y_\mathrm{tube} &=& \frac{r_\mathrm{tube}\sin\theta}{\sqrt{0.41\cos^2\theta+0.186\sin^2\theta}} \ \ \text{fm}. \nonumber
\end{eqnarray}
Here $r_\mathrm{tube}$ is used as a free parameter whose value is determined below, and the azimuthal angle of the tube $\theta$ is randomized among different events.

\begin{figure}
\begin{tabular}{cc}
\begin{minipage}{250pt}
\centerline{\includegraphics[width=250pt]{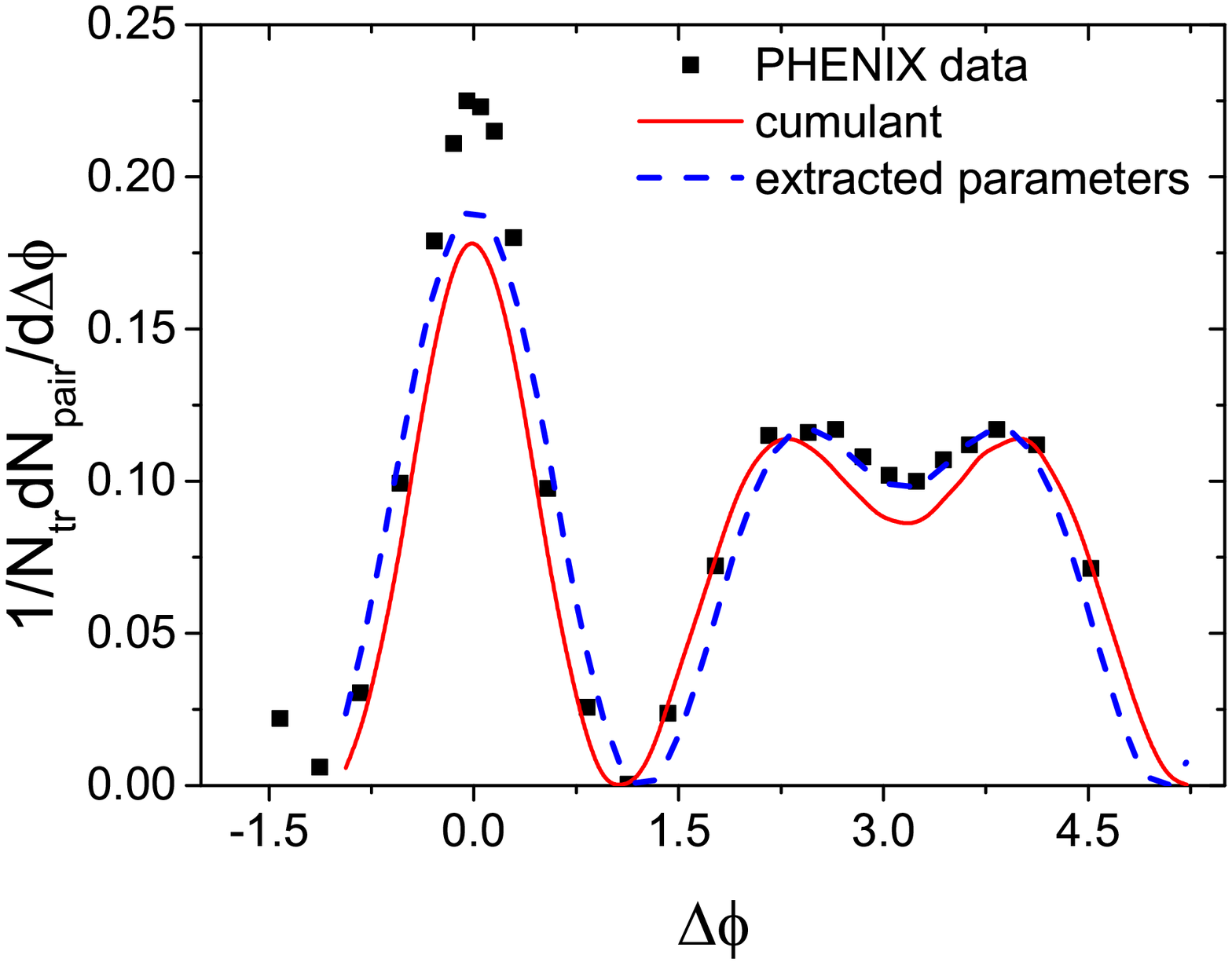}}
\end{minipage}
&
\begin{minipage}{250pt}
\centerline{\includegraphics[width=250pt]{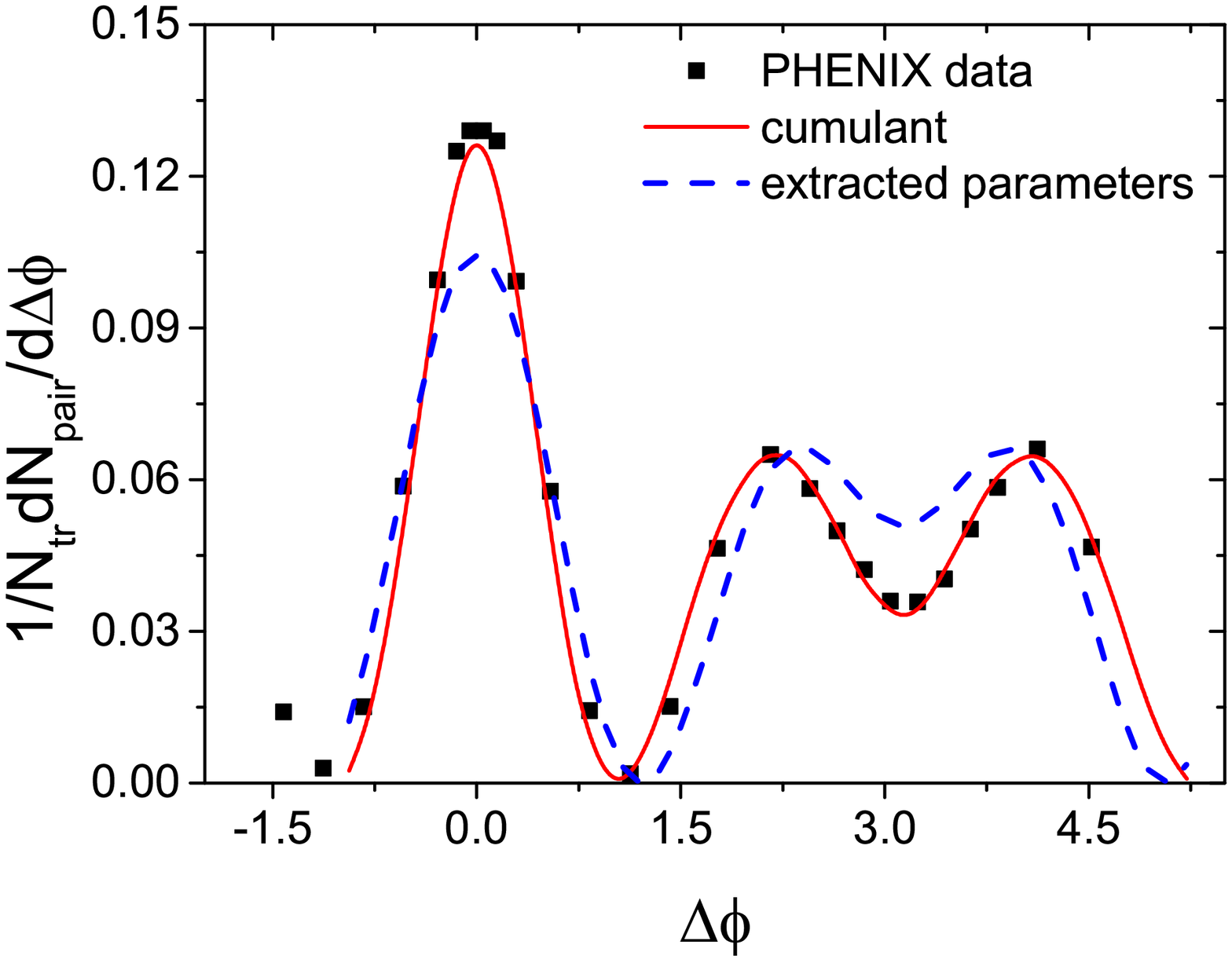}}
\end{minipage}
\end{tabular}
\caption{(Color online) The calculated two-particle correlations by using one-tube IC for $20\% - 40\%$ centrality window in comparison with the corresponding data by PHINEX Collaboration~\cite{RHIC-phenix-ridge-5}, and those obtained by using the extracted paramters in Table \ref{tubeparameters} and Eqs.(\ref{lbNbNt3}) and (\ref{lbNbNt4}).
The SPheRIO results from using cumulant method are shown in red solid curves, the data are shown in solid squares, and those obtained by the estimated parameters are shown by the blue hatched curves.
Left: the results for the momentum intervals $0.4<p_T^A<1$ Gev and $2<p_T^T<3$ Gev. 
Right: those for the momentum intervals $1<p_T^A<2$ Gev and $2<p_T^T<3$ Gev. }
 \label{fig2p}
\end{figure}

By combining the two pieces together, the IC for the present model read
\begin{eqnarray}
\epsilon = \epsilon_\mathrm{bgd}+\epsilon_\mathrm{tube} .
\end{eqnarray}
Subsequently, we carry out hydrodynamical simulations by using the SPheRIO code~\cite{sph-review-1,sph-corr-1,sph-corr-2,sph-corr-4,sph-eos-2,sph-cfo-1,sph-eos-3,sph-corr-ev-8}, it is an ideal hydrodynamic code using the Smoothed Particle Dynamics algorithm~\cite{sph-algorithm-review-01,sph-1st,sph-algorithm-09}.
For the present study, only the temporal evolution on the transverse plane is considered as Bjorken scaling is assumed.
We generate a total of 2000 events according to IC profile discussed above.
Then the hydrodynamical simulation is carried out for each event, by the end of which, a Monte-Carlo generator is evoked 200 times for hadronization in order to further increase the statistics.
The resultant two-particle correlations, evaluated by cumulant method, are shown in Fig.\ref{fig2p}, in comparison with the PHENIX data~\cite{RHIC-phenix-ridge-5}.
The paramter $r_\mathrm{tube}$ is chosen to be $2.3$ fm in the calculations.
As the hydrodynamic simulations are of two-dimension, the obtained correlations are multiplied by a factor related to the longitudinal scaling of the system.

Now, it is interesting to verify that the parameters given in Eq.(\ref{ccumulant2}) are indeed quantitatively meaningful.
To achieve this, we extract the model parameters in Eq.(\ref{ccumulant2}) using mostly the same arguments leading to very expression.
We first estimate those for the background distribution $\frac{dN_{\mathrm{bgd}}}{d\phi}$.
The background flow coefficients $v_2^\mathrm{bgd}$ can be obtained directly by investigating the hydrodynamic evolution of IC solely determined by $\epsilon_\mathrm{bgd}$.
A total of 2000 events with 200 Monte Carlo each are considered in the evaluation.
To estimate the multiplicity fluctuations of the background, $<N_\mathrm{bgd}^TN_\mathrm{bgd}^A>-<N_\mathrm{bgd}^T><N_\mathrm{bgd}^A>$, we count, on an event-by-event basis, the number of particles of corresponding momentum intervals. 
The events in question are those generated by NeXuS of 20\% - 40\% centrality window, and we made use of a total of 1000 events.
By using Fourier expansion of the two particle correlation, and extracting the second and third order coefficients, one obtains the parameters related to $v_2^{\mathrm{tube},T}$, $v_2^{\mathrm{tube},A}$, $v_3^{\mathrm{tube},T}$, and $v_3^{\mathrm{tube},A}$.
The obtained values are summarized in Table \ref{tubeparameters} and Eqs.(\ref{lbNbNt3}) and (\ref{lbNbNt4}).

\begin{table}[ht]
\begin{center}
\caption{The calculated background as well as overall elliptic flow coefficients for correspoding transverse momentum intervals of trigger and associated particles.
The calculations are carried out by using IC of smooth elliptic energy distribution as described in the text.}\vspace{0.5cm}
\begin{tabular}{c|c|c|c}
&  $0.4<p_T<1$ &  $1<p_T<2$ &  $2<p_T<3$\\
\hline
$v_{2}^\mathrm{bgd}$ &  0.11&    0.21&    0.36\\
\hline
$v_{2}^\mathrm{all}$ &  0.09&    0.17&    0.26\\
\end{tabular}
\label{tubeparameters}
\end{center}
\end{table}

On the other hand, however, some of the parameters can also be inferred straightforwardly from the experimental data.
Therefore, we carry out these comparisons with the published data as follow.
The overall elliptic flow, $v_2^\mathrm{all}$, obtained by simulations of event-by-event fluctuating IC consisting of the background and tube, should be consistent with the collisions of the same centrality window.
This is confirmed by comparing the value of $v_{2}^\mathrm{all}$ against the data of 20\%-60\% Au+Au collisions obtained by PHENIX~\cite{RHIC-phenix-v2-4}.
The multiplicity fluctuations estimated from simulations and for the corresponding momentum intervals are found to be
\begin{eqnarray}
\langle N^T_\mathrm{bgd}N^A_\mathrm{bgd}\rangle -\langle N^T_\mathrm{bgd}\rangle \langle N^A_\mathrm{bgd}\rangle =14.67    \, , \label{lbNbNt3} \\ 
\langle N^T_\mathrm{tube}N^A_\mathrm{tube}\rangle v_2^{\mathrm{tube},T}v_2^{\mathrm{tube},A}=1.62    , \nonumber\\
\langle N^T_\mathrm{tube}N^A_\mathrm{tube}\rangle v_3^{\mathrm{tube},T}v_3^{\mathrm{tube},A}=1.63    , \nonumber
\end{eqnarray}
for $0.4<p_T^A<1$, $2<p_T^T<3$, and
\begin{eqnarray}
\langle N^T_\mathrm{bgd}N^A_\mathrm{bgd}\rangle -\langle N^T_\mathrm{bgd}\rangle \langle N^A_\mathrm{bgd}\rangle =5.07  \, , \label{lbNbNt4}   \\ 
\langle N^T_\mathrm{tube}N^A_\mathrm{tube}\rangle v_2^{\mathrm{tube},T}v_2^{\mathrm{tube},A}=1.36    , \nonumber\\
\langle N^T_\mathrm{tube}N^A_\mathrm{tube}\rangle v_3^{\mathrm{tube},T}v_3^{\mathrm{tube},A}=1.48    , \nonumber
\end{eqnarray}
for $1<p_T^A<2$, $2<p_T^T<3$ respectively.
Finally, by substituting the above parameters back into Eq.(\ref{ccumulant2}), one obtains the two-particle correlation as also shown in Fig.\ref{fig2p}.
It is found that the two approaches are in good agreement with each other.

\begin{figure}
\begin{tabular}{ccc}
\vspace{0pt}
\begin{minipage}{150pt}
\centerline{\includegraphics[width=180pt]{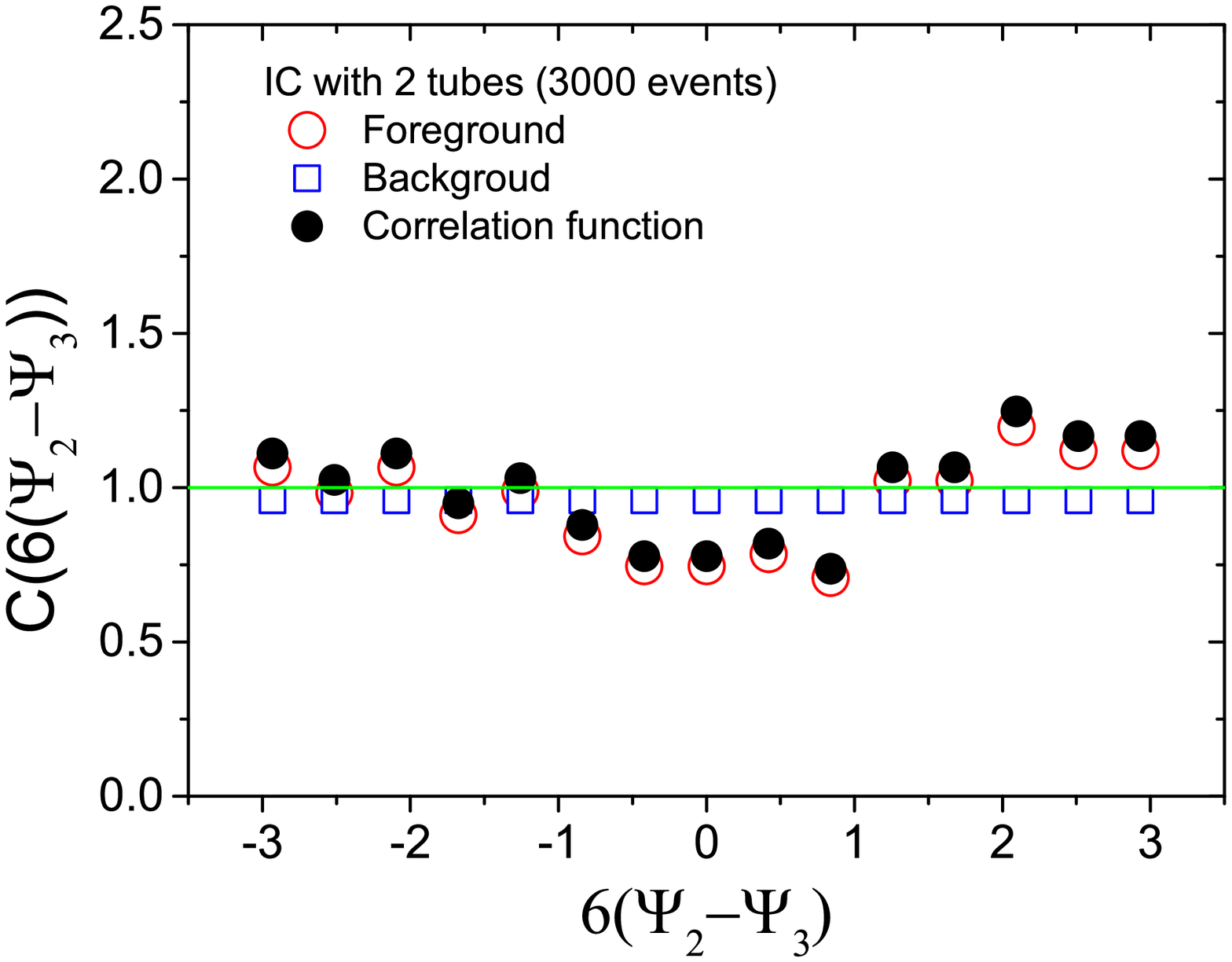}}
\end{minipage}
&
\begin{minipage}{150pt}
\centerline{\includegraphics[width=180pt]{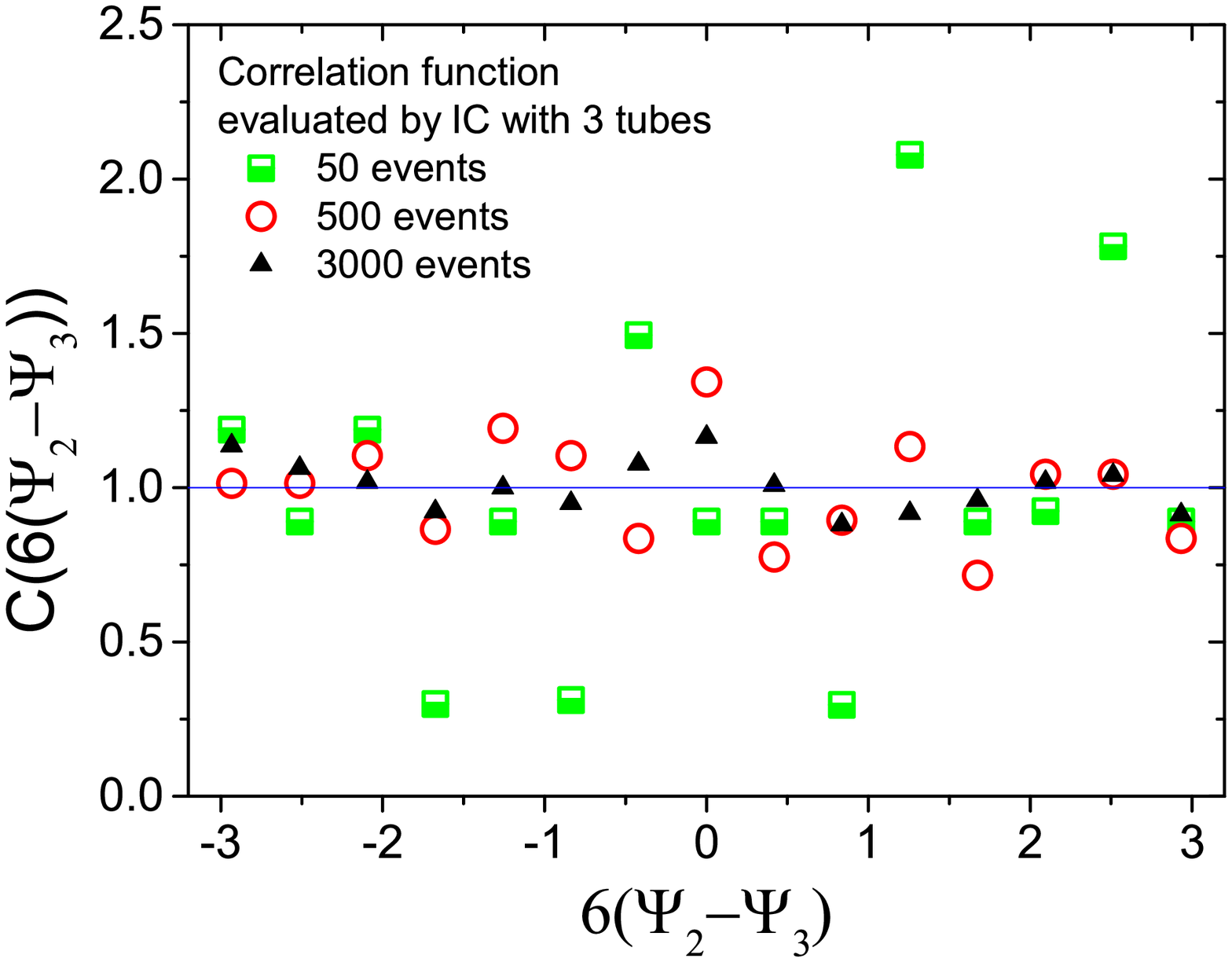}}
\end{minipage}
&
\begin{minipage}{150pt}
\centerline{\includegraphics[width=180pt]{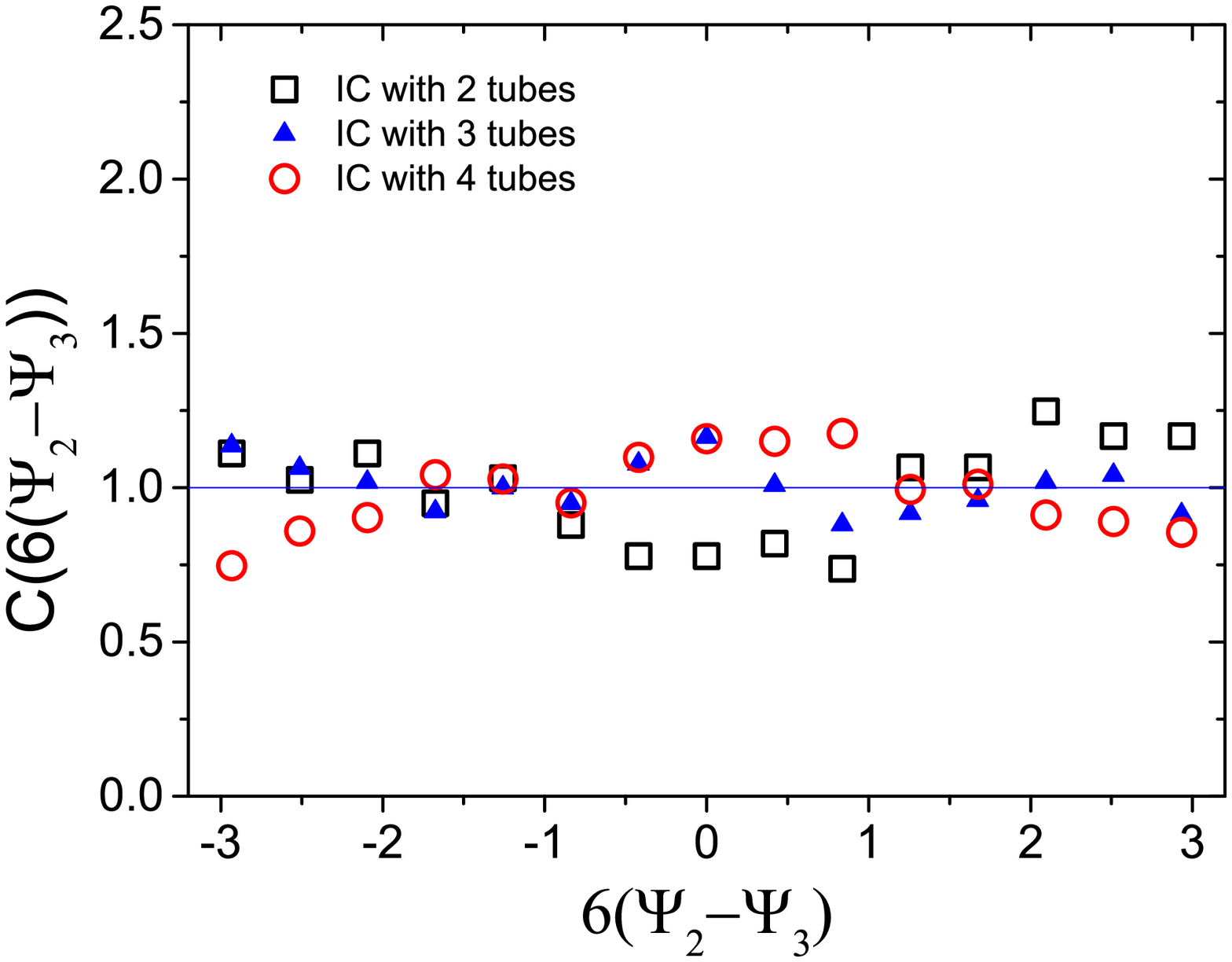}}
\end{minipage}
\\
\vspace{0pt}
\end{tabular}
\caption{(Color online) The relative angle distributions between $\Psi_2$ and $\Psi_3$ in tube model.
Left: The relative angle distribution (filled black circles) evaluated by the ratio of foreground (red empty circles) to the background (blue empty squares) signals. The results are for IC with two high energy tubes.
Middle: The results evaluated by using different numbers of IC with three high energy tubes.
Right: The relative angle distributions calculated by using IC with different number of tubes, where each curve is obtained by using 3000 events.}
 \label{figureCPsi23}
\end{figure}

Also, we investigated the correlation between different event planes.
In particular, since the elliptic flow and triangular flow from the tube are correlated by definition, it is interesting to verify whether the overall event planes $\Psi_2$ and $\Psi_3$ are correlated in comparison to existing data~\cite{LHC-atlas-vn-3}.
If one adds two high-energy tubes onto an isotropic background and assumes that the resultant distribution is just the superposition of those individual contributions, one observes that the resultant event planes might behave nonlinearly. 
By assuming that the azimuthal angles of the two tubes are $0$ and $\phi_t$, the harmonic coefficients are $v_n^{i}$ with $i=1, 2$, it is straightforward to find that
\begin{eqnarray} 
\Psi_2 = \frac12\arctan\frac{v_2^2\sin 2\phi_t}{v_2^1+v_2^2\cos 2\phi_t} \ ,\\
\Psi_3 = \frac13\arctan\frac{v_3^2\sin 3\phi_t}{v_3^1+v_3^2\cos 3\phi_t} \ .
\end{eqnarray}
Thus in general $\Psi_2 \ne \Psi_3$ if the harmonic coefficients of the two tubes are not precisely identical.
This has been demonstrated in our numerical study presented in Ref.~\cite{sph-corr-7}.
There the background is assumed to be isotropic, and all the tubes are identical, even though, numerical study shows that event planes are uncorrelated.
To confirm these results with better statistics, we carried out calculations in accordance with Ref.\cite{LHC-atlas-vn-3} by using a more significant number of events, where the relative angle distribution was obtained as the ratio of foreground to background signals.
Calculations were carried out by IC constructed using different numbers of tubes added to an isotropic background with randomized azimuthal angles.
As shown in Fig.\ref{figureCPsi23}, the relative angle distribution is found to approach to that of random distribution as the number of events increases. 
The feature is observed for IC with two, three and four high-energy tubes and found to be consistent with our previous findings.
It indicates that the tube model is consistent with the observed event plane correlation.

\section{IV. Concluding remarks}

The hydrodynamical approach has been shown to be a successful description of many experimental data of relativistic heavy ion collisions, prooving the emergence of collective flow dynamics. 
It is widely accepted that the flow dynamics are expressed by a set of flow harmonics, including correlations among them~\cite{hydro-corr-ph-2,hydro-corr-ph-4,hydro-corr-ph-6,hydro-corr-ph-7,hydro-corr-ph-8} on an event-by-event basis.  

On the other hand, if one wishes to investigate the properties of the matter such as the degree of local thermal equilibrium, namely, the equation of state and transport coefficients, we need to study observables sensitive to specific aspects of hydrodynamics which are genuinely nonlinear.  
Such behavior is, of course, contained in the intrinsically nonlinear, higher order correlators~\cite{hydro-corr-ph-7,hydro-corr-ph-8,hydro-vn-10}.
We understand that they may demonstrate themselves in an untrivial way and it will not be straightforward to extract a simple physical picture from them.  

Here, we note that several microscopic transport models, such as AMPT or PHSD, have shown to have similar properties as viscous hydrodynamic calculations~\cite{ampt-4,ampt-5,hydro-review-06}.
In particular, it is worth mentioning that in the event-by-event realizations, a state close to the local thermal equilibrium only corresponds to a tiny space-time domain during the entire dynamical evolution (Ref.~\cite{hydro-review-06,phsd-01}). 
To clarify up to what extent the genuine event-by-event hydrodynamics is valid, we should seek a new set of observables which are sensitive to the nonlinear evolution of the system.

The primary objective of the present paper is that, as a complementary effort to the current statistical analysis of flow harmonics, we propose a straightforward physical picture which describes the two-particle angular correlations.
It may serve as a possible candidate for the signals of the emergence of the local non-linear dynamics.  
In the standard approach, the double-peaked behavior is associated to the ocutople inhomogeneity in the initial energy distribution.  
Such a vision is relatively well-accepted, particularly when one considers the linear correlations between the event-averaged initial energy inhomogeneity $\langle \epsilon_n\rangle$ and the flow harmonics $\langle v_n\rangle$.
On the other hand, if we push the above arguments to the extreme, the concept of the ``local thermal equilibrium" becomes meaningless, since the correlations between $\langle v_3\rangle$ and $\langle \epsilon_3\rangle$ are only valid on the event-averaged basis.
Therefore, the nonlinear regime and fluctuations are of fundamental significance~\cite{hydro-corr-ph-7,hydro-corr-ph-8}.
It is true that the observed two-particle correlations suffer from the ambiguities of the subtraction process via, for example, the ZYAM method.  
However, once we establish a reasonable physically simple model, we can investigate the nonlinear properties and fluctuations within the framework of the standard flow harmonics analysis. 

Regarding the above points, we develop the peripheral tube model for the two-particle correlations and show that the data are consistent with those obtained by event-by-event hydrodynamical simulations using appropriately devised IC. 
The latter is further shown to be in qualitative agreement with those by using the analytic result of a simplified model with the parameters extracted from flow coefficients and multiplicity fluctuations. 

The model can be employed to study the event plane dependence of the two-particle correlaions~\cite{sph-corr-ev-4}, which can be derived as the modulation of the background evolves from in-plane to out-of-plane direction.
Also, the centrality dependence of the correlations has been discussed~\cite{sph-corr-ev-6}, where the model parameters were extracted from the NeXuS IC and were shown to give consistent results.
These results indicate that the peripheral tube model offers a simple physical vision for the interpretation of the observed features of the two-particle correlations in nuclear collisions. 

Note that the peripheral tube model, in general, generates the octupole anisotropy of the energy distribution, so that the correlations among $\epsilon_3$ and $v_3$ also appear. 
In this context, the analysis of fluctuations using higher order correlations may reveal the signal of the genuine nonlinear hydrodynamics. 
It is interesting to propose further and investigate observables which carry such signals. 
Quantities associated with the event plane correlations~\cite{LHC-atlas-vn-3,LHC-atlas-vn-5,LHC-alice-vn-5} and the 2+1 correlations~\cite{RHIC-star-corr-2plus1-1} may provide vital information. 
We believe that our approach, to be proven right or not, will offer a meaningful reference to investigate the nonlinear regime of the flow dynamics.
In this regard, many specific aspects of the model should be further clarified and explored in terms of standard flow harmonics. 
Works in this direction are under progress.

\section*{Acknowledgments}
We gratefully acknowledge the financial support from
Funda\c{c}\~ao de Amparo \`a Pesquisa do Estado de S\~ao Paulo (FAPESP),
Funda\c{c}\~ao de Amparo \`a Pesquisa do Estado do Rio de Janeiro (FAPERJ),
Conselho Nacional de Desenvolvimento Cient\'{\i}fico e Tecnol\'ogico (CNPq),
and Coordena\c{c}\~ao de Aperfei\c{c}oamento de Pessoal de N\'ivel Superior (CAPES).
A part of the work was developed under the project INCT-FNA Proc. No. 464898/2014-5, the Center for Scientific Computing (NCC/GridUNESP) of the S\~ao Paulo State University (UNESP),
and National Natural Science Foundation of China (NNSFC) under contract No.11805166.

\bibliographystyle{h-physrev}
\bibliography{references_qian}

\end{document}